\documentclass{article}

\PassOptionsToPackage{square,numbers,sort}{natbib} 

\usepackage[preprint]{main}

\usepackage[utf8]{inputenc} 
\usepackage[T1]{fontenc}    
\usepackage{hyperref}       
\usepackage{url}            
\usepackage{booktabs}       
\usepackage{amsfonts}       
\usepackage{nicefrac}       
\usepackage{microtype}      
\usepackage{xcolor}         
\usepackage{float}
\usepackage{amsmath}
\usepackage{multirow}
\usepackage{graphicx}
\usepackage{graphbox}
\usepackage{adjustbox}
\usepackage{caption}
\graphicspath{{Figures/}}

\title{HetSyn: Versatile Timescale Integration in Spiking Neural Networks via Heterogeneous Synapses}

\author{%
  Zhichao Deng\textsuperscript{1}\thanks{Equal contribution.}, ~Zhikun Liu\textsuperscript{1}\footnotemark[1], ~Junxue Wang\textsuperscript{1}, ~Shengqian Chen\textsuperscript{1}, ~Xiang Wei\textsuperscript{1}, ~Qiang Yu\textsuperscript{1,2}\thanks{Corresponding author.} \\
  \textsuperscript{1}College of Intelligence and Computing, Tianjin University, Tianjin, China\\
  \textsuperscript{2}College of Computer and Information Engineering, Tianjin Normal University, Tianjin, China \\
  \texttt{yuqiang@tju.edu.cn}
}

\begin{document}

\maketitle

\begin{abstract}
Spiking Neural Networks (SNNs) offer a biologically plausible and energy-efficient framework for temporal information processing. However, existing studies overlook a fundamental property widely observed in biological neurons—synaptic heterogeneity, which plays a crucial role in temporal processing and cognitive capabilities. To bridge this gap, we introduce HetSyn, a generalized framework that models synaptic heterogeneity with synapse-specific time constants. This design shifts temporal integration from the membrane potential to the synaptic current, enabling versatile timescale integration and allowing the model to capture diverse synaptic dynamics. We implement HetSyn as \mbox{HetSynLIF}, an extended form of the leaky integrate-and-fire (LIF) model equipped with \mbox{synapse-specific} decay dynamics. By adjusting the parameter configuration, \mbox{HetSynLIF} can be specialized into vanilla LIF neurons, neurons with threshold adaptation, and neuron-level heterogeneous models. We demonstrate that HetSynLIF not only improves the performance of SNNs across a variety of tasks—including pattern generation, delayed match-to-sample, speech recognition, and visual recognition—but also exhibits strong robustness to noise, enhanced working memory performance, efficiency under limited neuron resources, and generalization across timescales. In addition, analysis of the learned synaptic time constants reveals trends consistent with empirical observations in biological synapses. These findings underscore the significance of synaptic heterogeneity in enabling efficient neural computation, offering new insights into brain-inspired temporal modeling.

\end{abstract}

\section{Introduction}
\label{Introduction}


Spiking Neural Networks (SNNs) have been widely studied as a promising and energy-efficient alternative to conventional artificial neural networks (ANNs), offering a biologically plausible computing paradigm characterized by sparse, event-driven signaling and inherent capability for temporal processing~\cite{maass1997networks, pfeiffer2018deep, pei2019towards}. However, due to the complex dynamical characteristics of spiking neurons, effectively training SNNs and improving their performance remain major challenges in the field. This calls for new learning paradigms, either by adapting established methods from ANNs~\cite{wu2019direct, sengupta2019spikevgg, kim2020spikingyolo, zhou2023spikformer, yao2023spike, duan2022temporal, fang2021deep} or by drawing inspiration from biological mechanisms~\cite{brzosko2019neuromodulation, bellec2018long, bellec2020solution, liu2021celltype, zhang2023brain, wu2022brain}.

While ANN-derived methods have led to notable progress, we instead focus on strengthening the biological foundations of SNNs by revisiting fundamental neurobiological mechanisms, which remain underexplored and hold great potential for enabling more versatile timescale integration in spiking systems. This capability is particularly important, as effective timescale integration is central to temporal cognition, and many real-world tasks—such as working memory, speech recognition, and sequential decision-making—require neural systems to operate across multiple timescales~\cite{hasson2008hierarchy, buonomano2009state, kiebel2008hierarchy, baddeley2003working}. Achieving such multi-timescale processing requires neural models to incorporate diverse temporal dynamics across different input pathways. To this end, various forms of temporal heterogeneity have been introduced in SNNs to enhance their capability to represent and integrate information across multiple time constants, such as ALIF~\cite{bellec2018long}, PLIF~\cite{fang2021incorporating}, neuron heterogeneity~\cite{perez2021neural}, and dendritic heterogeneity~\cite{zheng2024temporal}. Although these approaches provide valuable insights, they remain limited to the neuron level or sub-neuronal level, where temporal dynamics are uniformly applied across aggregated inputs. This restricts their ability to assign distinct temporal responses to different synaptic pathways, limiting performance on tasks with complex temporal structure.

\begin{figure}[H]
  \centering
  \includegraphics[width=1.0\textwidth]{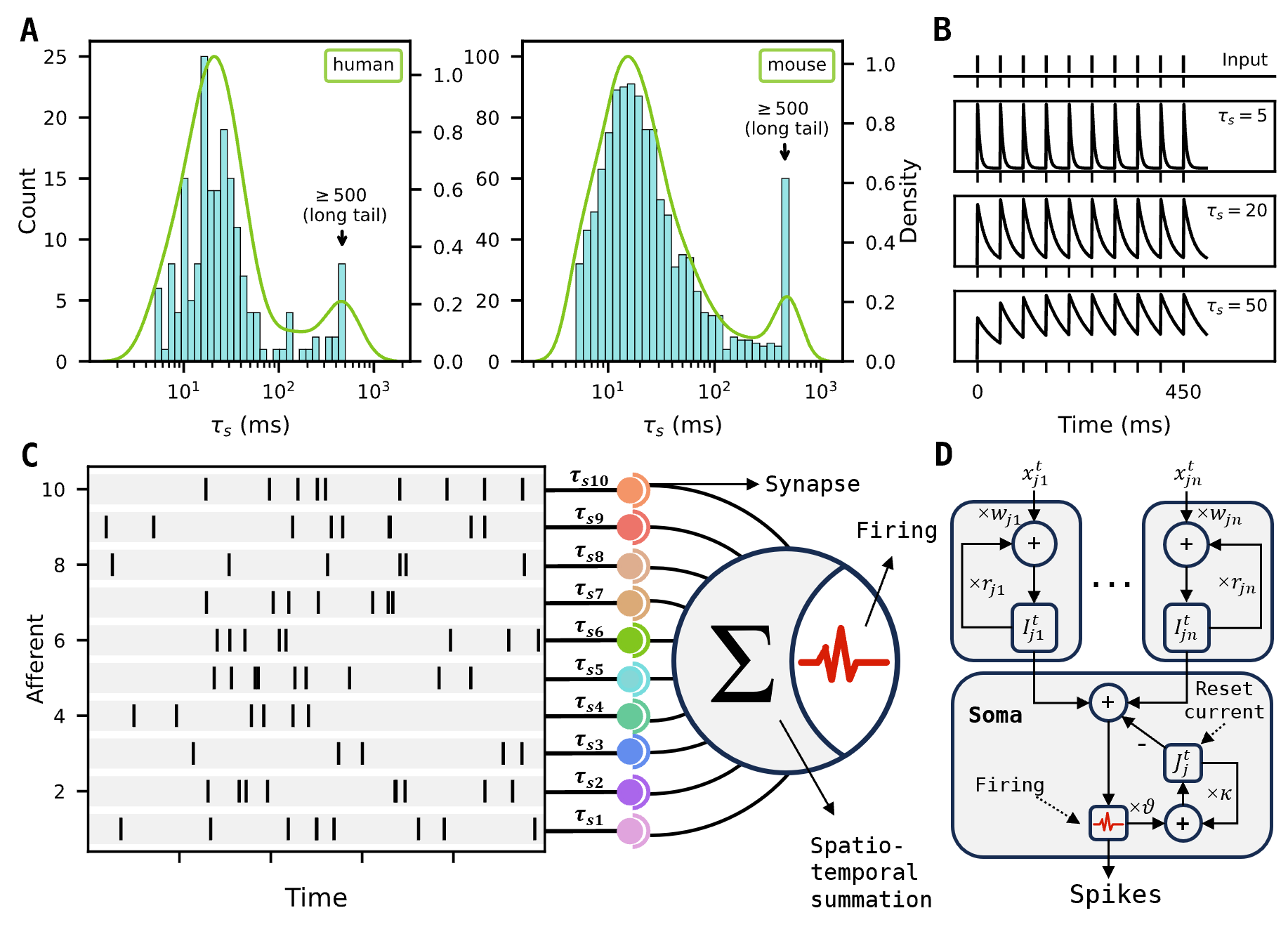}
  \caption{(A) Distributions of synaptic time constants ($\tau_\text{s}$) extracted from human (left, 194 pairs) and mouse (right, 1213 pairs) cortex, shown as histograms with overlaid KDE curves (green). The broad, long-tailed ($\tau_\text{s} \ge 500$ ms) profiles reflect substantial synaptic heterogeneity. (B) Postsynaptic membrane potential traces under different synaptic time constants ($\tau_\text{s}$ = 5, 20, 50~ms, from second to fourth row) in response to a regular spike train (first row). (C-D) Schematic of HetSyn, in which afferent spikes are integrated via synapses with heterogeneous decay dynamics.}
  \label{fig:fig1}
\end{figure}

In contrast, numerous neuroscience studies have shown that synapses vary substantially across brain regions and cell types~\cite{craig2001molecular, montgomery2002state}, giving rise to a diverse temporal basis that supports multi-timescale integration and cognitive abilities~\cite{ chabrol2015synaptic, bittner2017behavioral} (see Fig.~\ref{fig:fig1}A, B). This diversity, known as synaptic heterogeneity, constitutes a fundamental biological principle that plays a crucial role in shaping neural computation. Despite its biological significance, synaptic heterogeneity has rarely been incorporated into the design of spiking neural networks. Its computational potential remains largely unexplored, in part due to the challenges of modeling fine-grained, synapse-specific temporal dynamics.

To bridge this gap, we take a first step toward incorporating synaptic heterogeneity into SNNs and introduce HetSyn by focusing on one key aspect: the diversity of synaptic time constants, motivated by our analysis of a recently released Synaptic Physiology Dataset~\cite{campagnola2022local} from the Allen Institute, which reveals that synaptic time constants follow a log-normal distribution across populations, providing direct empirical support for synapse-specific time constants (Fig.~\ref{fig:fig1}A). As illustrated in Fig.~\ref{fig:fig1}B, synapses with longer time constants are able to retain input-driven activity over extended periods, whereas those with smaller time constants respond more transiently and can effectively filter out noise or irrelevant fluctuations. This functional diversity lays the foundation for versatile timescale integration, enabling neural models to process information across both long and short temporal windows.

Furthermore, instead of relying on a single membrane time constant, HetSyn computes the membrane potential by aggregating synaptic currents, each governed by a synapse-specific decay factor. This synapse-driven formulation allows temporal integration to arise from the diverse dynamics of individual synapses, rather than being uniformly controlled at the neuron level (Fig.~\ref{fig:fig1}C, D). We instantiate HetSyn as HetSynLIF and demonstrate that it subsumes several representative spiking neuron models as special cases through parameter configuration (see Methods). 

Our contributions are as follows: 

\begin{itemize}
    \item We propose HetSyn, the first modeling framework to explicitly explore synaptic heterogeneity in SNNs, offering a biologically plausible yet computationally powerful approach for versatile timescale integration.
    \item We demonstrate that HetSyn serves as a unified and extensible framework, capable of representing a wide range of existing spiking neuron models.
    \item We instantiate HetSyn as HetSynLIF and demonstrate its effectiveness across multiple temporal tasks, with consistently strong performance. Notably, we achieve 92.36\% accuracy on the SHD dataset, which is, to the best of our knowledge, the best-reported accuracy among models with similar network structures.
\end{itemize}

\section{Related Work}
\subsection{Training methods}
There are two primary approaches for training SNNs: ANN-to-SNN conversion (ANN2SNN)~\cite{cao2015spiking, han2020rmp, li2021free, bu2023optimal, zhou2024rethinking, hwang2024spike} and direct training using surrogate gradient methods~\cite{wu2018spatio, shrestha2018slayer, neftci2019surrogate, fang2021deep, xiao2021training, zhou2024qkformer}. The ANN2SNN method first trains a conventional ANN and then maps its parameters to an SNN, where the firing rates of spiking neurons are used to approximate the continuous activations of the original ANN. However, it often suffers from accuracy degradation due to the approximation process. In comparison, surrogate gradient methods approximate the non-differentiable spiking function with a surrogate, thus enabling direct end-to-end training of SNNs via backpropagation through time (BPTT). This approach forms the basis of modern surrogate-gradient based training frameworks for SNNs, such as Spatio-Temporal Backpropagation (STBP)~\cite{wu2018spatio} and Slayer~\cite{shrestha2018slayer}. In this paper, we adopt the surrogate gradient method to enable efficient training of the HetSynLIF model.

\subsection{Heterogeneity in SNNs}
Recent studies have highlighted the significance of heterogeneity in SNNs, exploring its impact on network performance and temporal processing from various perspectives. For instance, neuron-level heterogeneity has been shown to enhance the stability and noise robustness of SNNs. Perez-Nieves et al.~\cite{perez2021neural} show that networks with membrane time constants drawn from a Gamma distribution better capture complex temporal patterns, outperforming homogeneous networks in tasks with rich temporal structure. In addition, PLIF~\cite{fang2021incorporating} learns a shared membrane time constant for each layer, which can be interpreted as introducing heterogeneity at a coarser granularity. Another source of heterogeneity arises from the spiking history of a neuron~\cite{bellec2018long}. Firing thresholds that evolve over time based on recent spiking activity introduce a form of adaptive, history-dependent threshold heterogeneity. Furthermore, heterogeneity has also been explored at the level of neural dynamics and plasticity mechanisms. Chakraborty et al.~\cite{chakraborty2023heterogeneous} investigate the effects of heterogeneity in LIF and spike-timing-dependent plasticity (STDP) parameters. Additionally, the temporal heterogeneity of dendritic branches~\cite{zheng2024temporal} has been integrated into SNNs. While considerable work has explored heterogeneity in SNNs, synapse‐level heterogeneity remains largely underexplored. In contrast, our HetSyn introduces synapse-specific decay factors, embedding heterogeneity directly at the synaptic level. This fine-grained design enables HetSyn to integrate temporal information over a broader range of timescales, thereby enhancing both computational flexibility and biological plausibility.

\section{Methods}
\label{Methods}

\subsection{Vanilla LIF}
Spiking neurons are the fundamental computational units in SNNs, enabling the modeling of temporal dynamics through biologically inspired mechanisms. Among various neuron models, the LIF neuron is one of the most widely adopted for its balance between computational efficiency and biological plausibility. It captures essential neuronal properties such as membrane potential leakage, input integration, and spike emission. The dynamics of the LIF neuron can be formulated as follows:
\begin{align}
\frac{d V}{d t}=-\frac{V-V_{\text {rest}}}{\tau_{\text{m}}}+\sum_{i, j} w_{i} \cdot \delta\left(t-t_{i}^{j}\right)- \vartheta \cdot \sum_{j} \delta\left(t-t_{\text{s}}^{j}\right)
\label{eq:LIF_membrane_derivation}
\end{align}
where \(V\) is the membrane potential, \(V_{\text{rest}}\) is the resting potential, and \(\tau_{\text{m}}\) represents the membrane time constant. \(t_{i}^{j}\) and \(t_{\text{s}}^{j}\) denote the timing of the \(j\)-th input spike from the \(i\)-th presynaptic neuron and the \(j\)-th output spike from the postsynaptic neuron, respectively. The term \(w_{i}\) denotes the synaptic efficacy of the \(i\)-th afferent, \(\vartheta\) denotes the firing threshold and \(\delta(\cdot)\) represents the Dirac delta function. We set \(V_{\text{rest}}=0\) in this paper, and an equivalent continuous-time solution to Eq.~(\ref{eq:LIF_membrane_derivation}) is then given by:
\begin{align}
V^{t} = \sum_{i} w_{i} \sum_{j} k\left(t - t_{i}^{j}\right) - \vartheta \cdot \sum_{j} k\left(t - t_{\text{s}}^{j}\right)
\label{eq:LIF_membrane}
\end{align}
where \(V^{t}\) is the membrane potential at time \(t\). The term \(k\left(t - t_{i}^{j}\right) = \exp({- \frac{t - t_{i}^{j}}{\tau_{\text{m}}}})\), for \( t > t_i^j \), is the synaptic kernel induced by the \(j\)-th spike of the \(i\)-th afferent. It describes the resulting postsynaptic potential (PSP), which decays exponentially at a rate governed by the membrane time constant \(\tau_{\text{m}}\). In practice, simulations typically adopt a discrete-time formulation, which is given by:
\begin{align}
V^{t}&=\rho \cdot V^{t-1}+\sum_{i} w_{i} \cdot z_{i}^{t}- \vartheta \cdot z^{t-1}
\label{eq:LIF_membrane_discrete} \\
z^{t}&=\text{H}(V^{t} -\vartheta)
\end{align}
where \(\rho=\exp(-\frac{\Delta t}{\tau_{\text{m}}})\) denotes the membrane decay factor, with \(\Delta t\) representing the discrete timestep. The output spike \(z^{t}\) is computed using the Heaviside step function \(\text{H}(\cdot)\).

\subsection{LIF-based spiking neuron with HetSyn}

We apply the HetSyn principle to the vanilla LIF neuron and propose HetSynLIF, a variant that incorporates synaptic heterogeneity. In HetSynLIF, the traditional membrane time constant is replaced by synapse-specific time constants. Instead of applying a uniform temporal filter to all inputs, each synaptic input is individually integrated through its own exponential decay, resulting in heterogeneous synaptic currents. This design shifts temporal integration from the membrane dynamics to the synaptic dynamics, reproducing synaptic diversity in biological systems. Moreover, we model the reset mechanism as a negative current injection. The neuron then accumulates these individually filtered synaptic currents and subtracts the reset current to update its membrane potential. The dynamics of HetSynLIF can be derived from Eq.~(\ref{eq:LIF_membrane}), yielding:
\begin{align}
V^{t} &= \sum_{i} I_{i}^{t} - J^{t}
\label{eq:HetSyn_membrane} \\
\frac{d I_i^{t}}{d t} &=-\frac{I_i^{t}}{\tau_{\text{s}, i}}+\sum_{j} w_{i} \cdot \delta\left(t-t_{i}^{j}\right) \\ 
\frac{d J^{t}}{d t} &=-\frac{J^{t}}{\tau_{J}}+\sum_{j} \vartheta \cdot \delta\left(t-t_{\text{s}}^{j}\right)
\end{align}
where \(I_i^{t}\) denotes the synaptic current from the \(i\)-th afferent at time \(t\), governed by its own synapse-specific time constant \(\tau_{\text{s}, i}\), thereby capturing the heterogeneity in synaptic temporal dynamics. \(J^{t}\) denotes the reset current, a neuron-level term that models the effect of spike-triggered reset and decays with time constant \(\tau_{J}\). The discrete-time formulations of synaptic and reset currents are given by:
\begin{align}
I_{i}^{t}&=r_{i} \cdot I_{i}^{t-1}+w_{i} \cdot z_{i}^{t}
\label{eq:HetSyn_current} \\
J^{t}&=\kappa \cdot J^{t-1}+\vartheta \cdot z^{t-1}
\label{eq:HetSyn_spike}
\end{align}
where \(r_{i}=\exp(-\frac{\Delta t}{\tau_{\text{s}, i}})\) and \(\kappa=\exp(-\frac{\Delta t}{\tau_{J}})\) are the decay factors for the \(i\)-th synaptic current and the reset current of the neuron, respectively.

\subsection{Generalize to other spiking neurons}
\label{sec:generalize}

Equipped with HetSyn, our HetSynLIF model is highly flexible and can be generalized to a wide range of spiking neuron models. To distinguish between variants, we use the prefixes "HomNeu" and "HetNeu" to denote homogeneous and heterogeneous configurations at the neuron level, respectively. Under this notation, HetSynLIF can reduce to vanilla LIF neurons (HomNeuLIF), neurons with threshold adaptation (HomNeuALIF), and neuron-level heterogeneous models (HetNeuLIF). This flexibility stems from the incorporation of synapse-specific time constants, which enable HetSynLIF to modulate temporal dynamics at a finer granularity. By analyzing the synaptic dynamics from a presynaptic neuron \(i\) to a postsynaptic neuron \(j\), we demonstrate how HetSynLIF generalizes to other neuron models under specific conditions. See Appendix~\ref{Theoretical Proof} for detailed proof.

\paragraph{Proposition 1} {\itshape If all synaptic decay factors \(r_{ji}\) and the reset current decay factor \(\kappa_j\) are identical and equal to a shared value \(\rho\), i.e., \(r_{ji} = \kappa_j = \rho\) for all \(i, j\), then the HetSynLIF model reduces to the HomNeuLIF model. Under this condition, the membrane potential of the postsynaptic neuron \(j\) at time \(t\) in HetSynLIF simplifies to:}
\begin{align}
V_{j}^{t} = \rho \cdot V_{j}^{t-1} + \sum_{i} w_{ji} \cdot z_{i}^{t} - \vartheta \cdot z_{j}^{t-1}
\end{align}
The resulting equation matches Eq.~(\ref{eq:LIF_membrane_discrete}), which suggests that the vanilla LIF model emerges as a special case of HetSynLIF when all synaptic and reset decay factors are identical.

\paragraph{Proposition 2} \textit{Based on Proposition~1, if we further decompose the reset current into a standard component and an additional adaptation current triggered by spikes—characterized by a decay factor \( \rho_{a} \) and a scaling coefficient \( a \)—then HetSynLIF generalizes to the HomNeuALIF model. Under this condition, the membrane potential of postsynaptic neuron \(j\) at time \(t\) in HetSynLIF can be transformed into:}
\begin{align}
V_{j}^{t} =\rho \cdot V_j^{t-1} + \sum_i w_{ji} \cdot z_i^t - J_{a}^{t}-\vartheta \cdot z_{j}^{t-1}
\end{align}
This indicates that HetSynLIF effectively reproduces the adaptive dynamics of the HomNeuALIF model proposed by \cite{rao2022long} through a structural modification of the reset current, demonstrating its flexibility in modeling neuron dynamics beyond fixed-threshold designs.

\paragraph{Proposition 3} \textit{For each postsynaptic neuron \(j\), if all synaptic decay factors \(r_{ji}\) and the reset current decay factor \(\kappa_j\) are identical and equal to a neuron-specific membrane decay factor \(\rho_j\), i.e., \(r_{ji} = \kappa_j = \rho_j\) for all \(i\), then HetSynLIF generalizes to the HetNeuLIF model. Under this condition, the membrane potential of postsynaptic neuron \(j\) at time \(t\) in HetSynLIF simplifies to:}
\begin{align}
V_j^t = \rho_j \cdot V_j^{t-1} + \sum_i w_{ji} \cdot z_i^t - \vartheta \cdot z_j^{t-1}
\end{align}
In this case, each postsynaptic neuron \(j\) evolves according to its own membrane decay factor \(\rho_j\), which reflects neuron-level heterogeneity\cite{perez2021neural}. This shows that HetSynLIF encompasses neuron-level heterogeneity as a special case by assigning shared decay dynamics per neuron.

\section{Experiments}
As detailed in Section~\ref{sec:generalize}, HetSynLIF can be reduced to HomNeuLIF, HomNeuALIF, and HetNeuLIF through appropriate parameterization. In this section, we conduct a systematic comparison of the four variants and further evaluate HetSynLIF against previous state-of-the-art methods. All variants are implemented within our unified framework to ensure theoretical equivalence, controlled ablation, fair comparison, and architectural consistency, enabling us to isolate the specific contribution of synapse-level heterogeneity. For clarity, we prefix feedforward and recurrent SNN implementations with "F-" and "R-", respectively. Further experimental and training details are provided in the Appendix.


\subsection{Versatile Timescale Integration in Pattern Generation Task}


\begin{figure}[H]
  \includegraphics[width=1.0\textwidth]{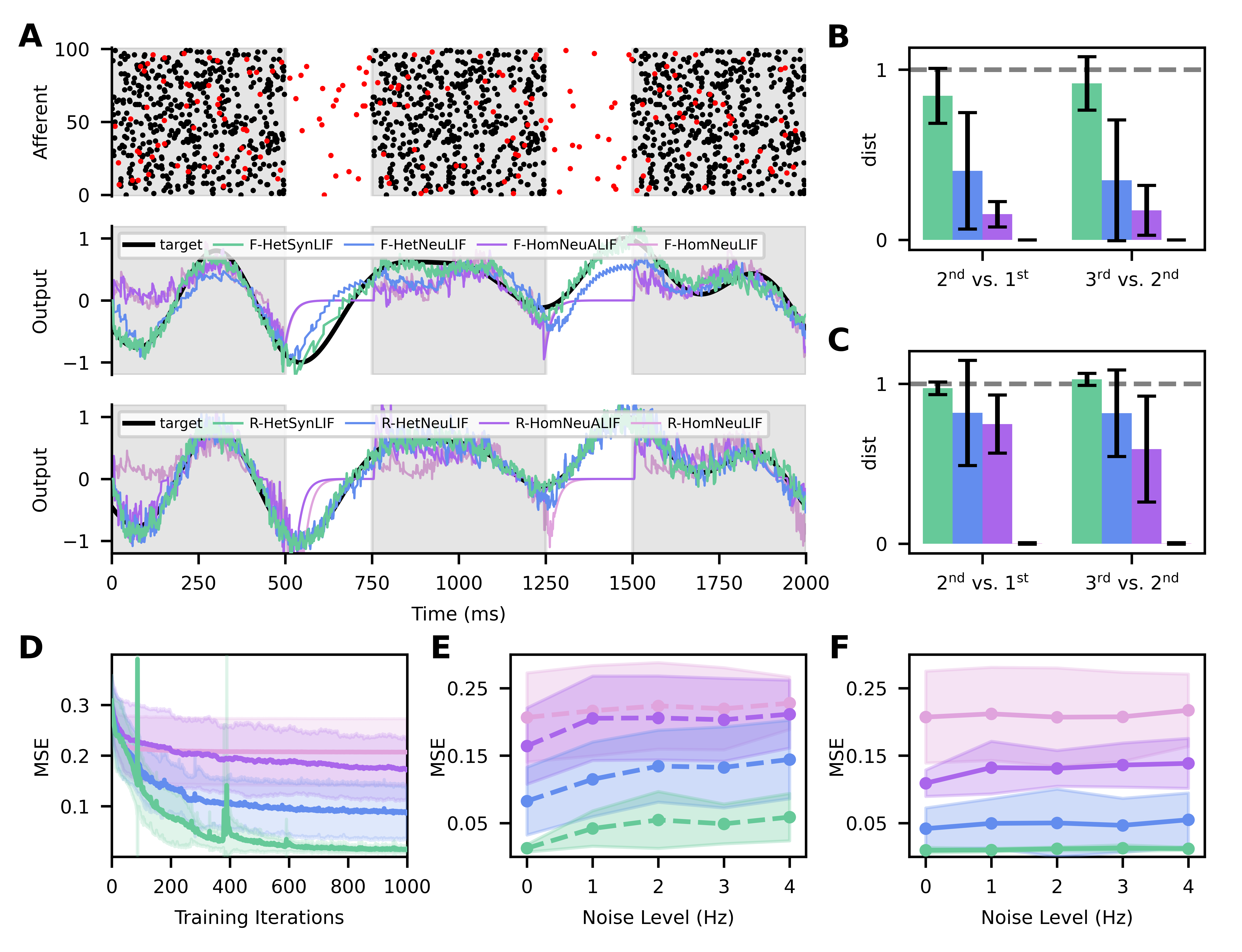}
  \caption{Pattern generation task. (A) An input spike pattern with a fixed Poisson-generated template (10 Hz, black dots) and superimposed noise spikes (2 Hz, red dots). Repeating features are marked by shaded gray regions (Top). Output traces of four FSNN variants (middle) and four RSNN variants (bottom) in response to the same input pattern. (B-C) Normalized distances between output traces at the \( 2^{\text{nd}} \) vs. \( 1^{\text{st}} \) and \( 3^{\text{rd}} \) vs. \( 2^{\text{nd}} \) occurrences of the repeating input segments for FSNNs (B) and RSNNs (C). Higher means better; see Appendix \ref{Details in Patter Generation task}. (D) Mean squared error over training iterations for four FSNN variants. (E-F) Mean squared error of FSNN variants (E) and RSNN variants (F) under different noise levels. Data in \textbf{B}-\textbf{F} are averaged over 10 runs and reported as mean $\pm$ s.d.
}
  \label{fig:fig2}
\end{figure}
We first evaluate HetSynLIF and three variants using one-layer FSNN and RSNN architectures on a more complex version of the pattern generation task, compared to the setups used in previous works~\cite{bellec2020solution, liu2021celltype}, where the network is trained to reproduce a target trace in response to structured input spike patterns that consist of repeated segments embedded in Poisson noise, with no input provided between the segments (Fig.~\ref{fig:fig2}A; see Appendix for details). We reveal that our HetSynLIF exhibits the fastest convergence and lowest mean squared error (MSE). In contrast, both HomNeuLIF and HomNeuALIF produce near-zero outputs during the input-free intervals between repeated segments, regardless of whether the network is an FSNN or an RSNN. This suggests their limited ability to learn time constants long enough to bridge these temporal gaps (Fig.~\ref{fig:fig2}A, D). We then compute normalized distances between outputs corresponding to repeated input segments (Fig.~\ref{fig:fig2}B for FSNNs, Fig.~\ref{fig:fig2}C for RSNNs, see Appendix \ref{Details in Patter Generation task}), where larger distances indicate a stronger ability to generate target-aligned outputs for identical input segments. HomNeuLIF shows consistently low distances, indicating nearly identical responses, while HomNeuALIF and HetNeuLIF show slight improvements in FSNNs and modest gains in RSNNs, benefiting from recurrent connections. Nevertheless, all three are inferior to HetSynLIF in performance, which consistently achieves the highest distances across both architectures. Additionally, under varying levels of input noise, HetSynLIF maintains the lowest MSE, while other models exhibit greater degradation (Fig.~\ref{fig:fig2}E, F). Finally, we show that HetSynLIF is also capable of generating multiple target patterns simultaneously (Appendix Fig.~\ref{fig:fig5}).

\subsection{Versatile Timescale Integration in Delayed Match-to-Sample Task}

\begin{figure}[H]
  \centering
  \includegraphics[width=1.0\textwidth]{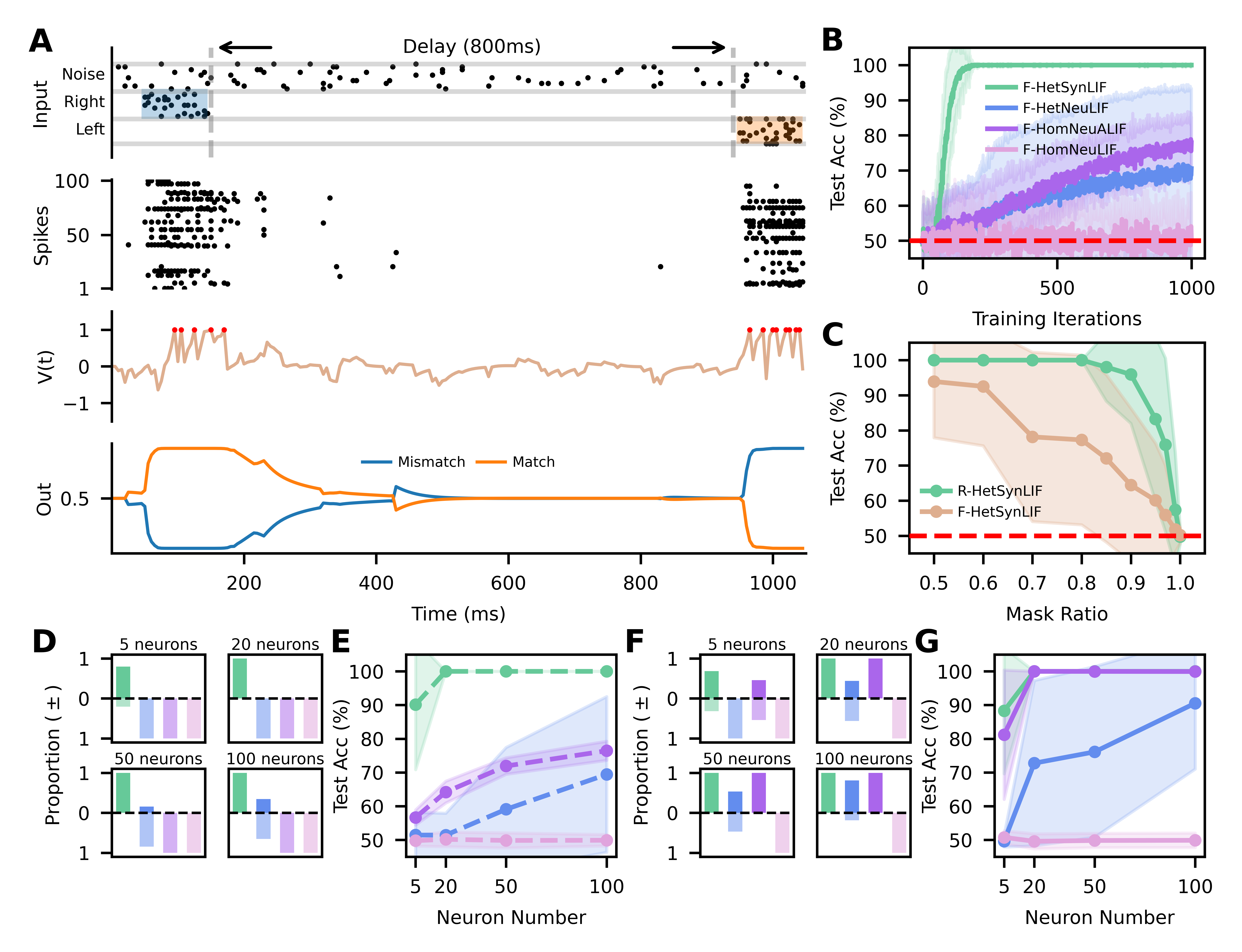}
  \caption{Delayed match-to-sample task. (A) Schematic diagram of a sample mismatch trial with two cues separated by an 800 ms delay (Top). Spiking activity of hidden layer neurons (Second). Membrane potential and spike activities (red dots) of a sample neuron (Third). Softmax outputs of neurons representing match and mismatch (Bottom). (B) Learning curves of four FSNN variants with 100 hidden neurons. (C) Task accuracy under different synaptic time constant masking ratios. (D) Success and failure rates across neuron numbers and FSNN variants. Each grid cell corresponds to a different hidden neuron count (5, 20, 50, 100), containing four centered bars for the four FSNN variants. Each bar is split vertically into a success segment (opaque, above) and a failure segment (transparent, below); trials with accuracy $\geq$ 90\% are considered successful. \textbf{D}-\textbf{G} follow the same color scheme as \textbf{B}. (E) Overall task accuracy of four FSNN variants under varying neuron counts. (F-G) Same as \textbf{D}-\textbf{E}, but for RSNN variants. Data in \textbf{B}-\textbf{G} are averaged over 50 runs and reported as mean $\pm$ s.d.}
  \label{fig:fig3}
\end{figure}

To assess the ability of HetSynLIF to capture multi-timescale dependencies essential for working memory, we adopt the delayed match-to-sample task—a widely used paradigm in neuroscience~\cite{miller2001integrative, romo2003flutter, chaudhuri2015large, rungratsameetaweemana2025random}. The network must judge whether two temporally separated cues belong to the same category (i.e., left or right), requiring short-term processing at cue onset and long-term retention across the delay (Fig.~\ref{fig:fig3}A; see Appendix for details). As in the pattern generation task, we compare the performance of eight variants—four FSNNs and four RSNNs—across multiple metrics. Fig.~\ref{fig:fig3}B presents the learning curves of the four FSNN variants, showing that F-HetSynLIF converges fastest and achieves the highest accuracy among FSNN variants. To assess the functional importance of synaptic heterogeneity, we progressively mask a fraction of the synaptic time constants in R-HetSynLIF, rendering them non-trainable. Results show that even with 80\% of synapses masked, the model maintains 100\% accuracy, highlighting the critical role of heterogeneous temporal integration in sustaining memory over delays (Fig.~\ref{fig:fig3}C). We further analyze how performance scales with network size (Fig. \ref{fig:fig3}D–G). Both F-HetSynLIF and R-HetSynLIF consistently achieve higher success rates (defined as accuracy $\geq$ 90\%) across varying numbers of hidden neurons, remaining effective even with five neurons, which highlights computational efficiency of synaptic heterogeneity in resource-constrained scenarios. Besides, R-HetSynLIF maintains 100\% accuracy even at a 2500~ms delay, with no notable increase in iterations to reach 95\% accuracy, and retains nearly 80\% accuracy under noise of 20~Hz, demonstrating both stable training efficiency and strong robustness (Appendix Fig.~\ref{fig:fig6}).

\subsection{Versatile Timescale Integration in Speech Recognition}
\label{Speech Recognition Section}

\begin{figure}[H]
  \includegraphics[width=1.0\textwidth]{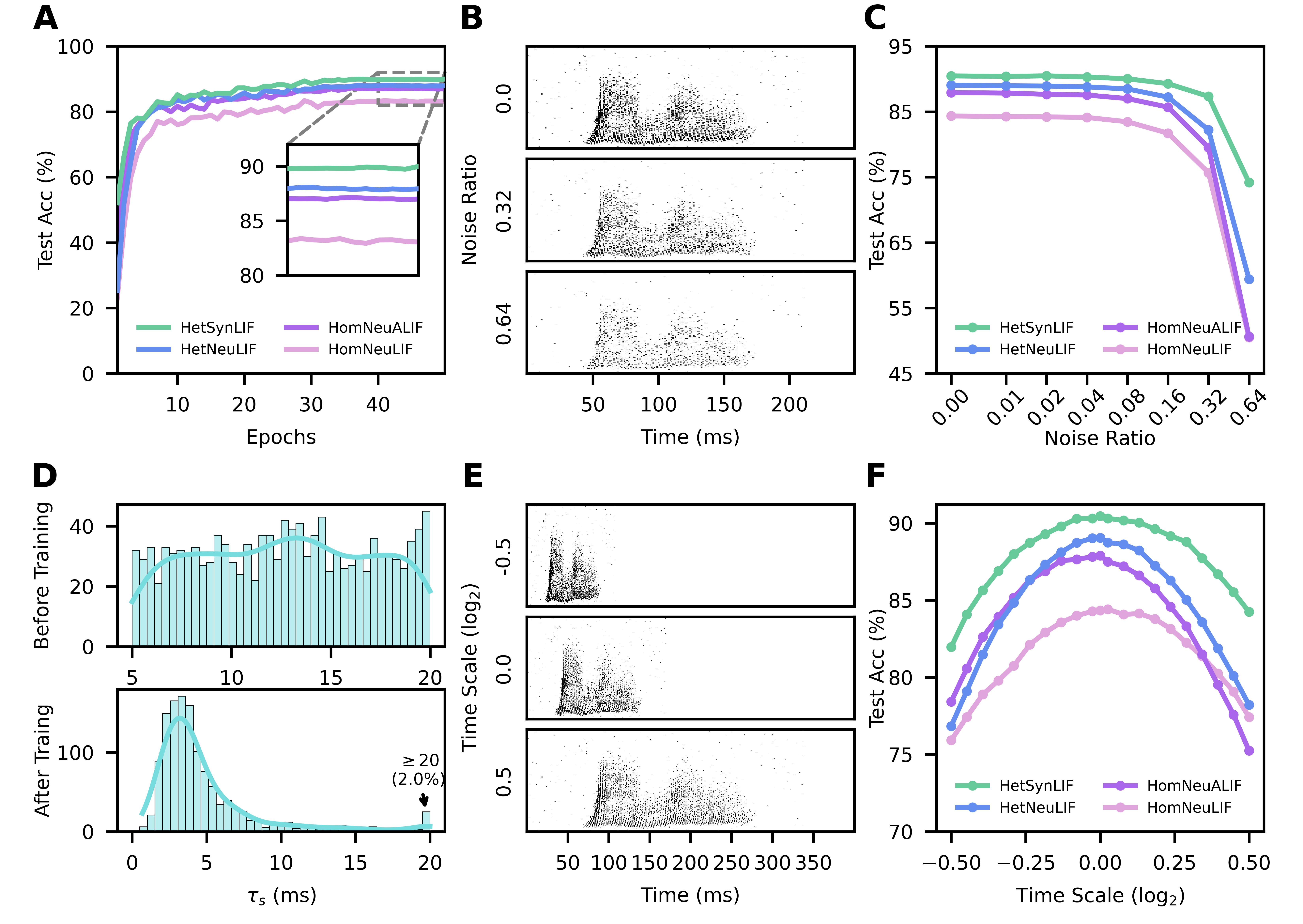}
  \caption{Speech recognition and robustness analysis. (A) Learning curves of four RSNN variants on the SHD dataset. (B) Spike inputs under deletion noise ratios of 0.00, 0.32, and 0.64 (top to bottom), where each ratio denotes the probability of deleting a spike. (C) Test accuracy under varying deletion noise ratios. (D) Distribution of synaptic time constants for connections from the last hidden layer to the output layer, before (top) and after (bottom) training. (E) Spike inputs under time warp conditions with $\text{log}_2$-scaled factors of -0.5 (compressed), 0.0 (original), and 0.5 (stretched), top to bottom. (F) Test accuracy under $\text{log}_2$-scaled time warp factors. Data in \textbf{A}, \textbf{C}, and \textbf{F} are averaged over 10 runs.}
  \label{fig:fig4}
\end{figure}

We then evaluate the speech recognition capability of four RSNN variants mentioned above on the SHD\cite{cramer2020heidelberg} dataset, using a two-layer recurrent architecture. As shown in Fig.~\ref{fig:fig4}A, HetSynLIF consistently achieves the highest accuracy and fastest convergence, followed by HetNeuLIF, HomNeuALIF, and HomNeuLIF. This result underscores the superior modeling capacity of synapse heterogeneity. While neuron heterogeneity also improves performance, assigning distinct time constants to individual synapses offers finer temporal resolution. In contrast, threshold adaptation yields limited gains and remains less effective than neuron heterogeneity. To understand how temporal dynamics evolve during learning, we analyze the distribution of $\tau_\text{s}$ in the final hidden-to-output layer (Fig.~\ref{fig:fig4}D). Initialized from \( \mathcal{U}(5, 20) \)~ms, $\tau_\text{s}$ shifts toward lower values, with 2.0\% exceeding 20 ms, forming a mild long-tail that may support slow-varying temporal encoding, loosely resembling biological patterns in Fig.~\ref{fig:fig1}A. Similar trends are observed across other layers (Appendix Fig.~\ref{fig:fig8}). We further assess robustness under deletion noise and time warp, both of which are introduced only during testing (i.e., trained on clean data and tested with deletion noise or time-warped inputs). In deletion noise, each spike is independently removed with a fixed probability (Fig.~\ref{fig:fig4}B). Although performance degrades with increasing noise, the accuracy ranking remains unchanged (Fig.~\ref{fig:fig4}C). Notably, when the noise ratio increases from 0.32 to 0.64, HetSynLIF maintains relatively stable, while others drop sharply, demonstrating the strong resilience of synapse heterogeneity to input disruptions. A similar trend holds under time warp distortions (Fig.~\ref{fig:fig4}E–F), where HetSynLIF consistently achieves higher accuracy across varying timescales (see Appendix~\ref{Details in Speech Recognition Task} and Fig.~\ref{fig:fig7} for more details).

\subsection{Comparison with Existing Works}
\label{Comparison with Existing Works}

We compare our proposed HetSynLIF model with other existing works on the SHD, S-MNIST~\cite{deng2012mnist}, TiDigits~\cite{leonard1993tidigits}, and Ti46-Alpha~\cite{liberman1993ti46} datasets, and report the results in Table~\ref{tab:accuracy}. Notably, our two-layer RSNN architecture achieves 92.36\% accuracy on the SHD dataset, which, to the best of our knowledge, is the highest reported accuracy among models with similar architectures. In addition, the HetSynLIF model demonstrates outstanding performance on the other three datasets as well, consistently outperforming prior methods, highlighting its effectiveness in processing multi-timescale temporal dynamics across both speech and visual recognition tasks.

\begin{table}[H]
        \caption{Accuracy comparison on SHD, S-MNIST, TiDigits, and Ti46-Alpha datasets}
        \vspace{0.5 \baselineskip}
        \label{tab:accuracy}
        \centering
        \begin{tabular}{llllll}
            \toprule
                Dataset & Method & Acc & Dataset & Method & Acc \\
            \cmidrule(r){1-6}
    \multirow{8}{*}{SHD}&$\mathrm{DH}$-$\mathrm{SFNN}^{\mathrm{\small{Nat.Commun24}}}$\cite{zheng2024temporal}&92.1&\multirow{8}{*}{TiDigits}&$\mathrm{BPT}$-$\mathrm{SNN}^{\mathrm{}}$\cite{lin2023bipolar}&98.1\\
    &$\mathrm{DH}$-$\mathrm{SRNN}^{\mathrm{Nat.Commun24}}$\cite{zheng2024temporal}&91.34&&$\mathrm{M}$-$\mathrm{STIP}^{\mathrm{TNNLS22}}$\cite{luo2022supervised}&98.1\\
    &$\mathrm{SRNN}^{\mathrm{ICLR25}}$\cite{ding2025rethinking}&91.19&&$\mathrm{MPD}$-$\mathrm{AL}^{\mathrm{AAAI19}}$\cite{zhang2019mpd}&97.5\\
    &$\mathrm{SRNN}^{\mathrm{NMI21}}$\cite{yin2021accurate}&90.4&&$\mathrm{BAE}$-$\mathrm{MPDAL}$\cite{pan2020efficient}&97.4\\
    &$\mathrm{NeuHet}$-$\mathrm{SRNN}^{\mathrm{NatCom21}}$\cite{perez2021neural}&82.7&&$\mathrm{PBSNLR}$-$\mathrm{DW}$\cite{zhang2020supervised}&96.5\\
    &$\mathrm{SRNN}^{\mathrm{PNAS22}}$\cite{cramer2022surrogate}&81.6&&$\mathrm{SpikingCNN}$\cite{tavanaei2017bio}&96.0\\
            \cmidrule(r){2-3}
            \cmidrule(r){5-6}
    &\textbf{HetSynLIF (Ours)}&\textbf{92.36}&&\textbf{HetSynLIF (Ours)}&\textbf{98.99}\\
            \cmidrule(r){1-6}
    \multirow{7}{*}{S-MNIST}&$\mathrm{DH}$-$\mathrm{SRNN}^{\mathrm{Nat.Commun24}}$\cite{zheng2024temporal}&98.87&\multirow{7}{*}{Ti46}&$\mathrm{RSNN}^{\mathrm{}}$\cite{zhang2024composing}&96.44\\
    &$\mathrm{SRNN}^{\mathrm{NMI21}}$\cite{yin2021accurate}&98.7&&$\mathrm{ScSr}$-$\mathrm{SNNs}$\cite{zhang2021skip}&95.9\\
    &$\mathrm{LSTM}^{\mathrm{ICML16}}$\cite{arjovsky2016unitary}&98.2&&$\mathrm{RSNN}^{\mathrm{NeurIPS19}}$\cite{zhang2019spike}&93.5\\
    &$\mathrm{LSNN}^{\mathrm{NeurIPS18}}$\cite{bellec2018long}&96.4&&$\mathrm{LSM}^{\mathrm{}}$\cite{zhang2015digital}&92.3\\
    &$\mathrm{AHP}$-$\mathrm{SNN}^{\mathrm{NMI22}}$\cite{rao2022long}&96.0&&$\mathrm{S}$-$\mathrm{MLP}^{\mathrm{NeurIPS18}}$\cite{jin2018hybrid}&90.98\\
            \cmidrule(r){5-6}
            \cmidrule(r){2-3}
    &\textbf{HetSynLIF (Ours)}&\textbf{98.93}&&\textbf{HetSynLIF (Ours)}&\textbf{96.53}\\
            \bottomrule
        \end{tabular}
        \label{table:tb1}
    \end{table}

\section{Conclusion}
\label{Conclusion}

In this paper, we propose HetSyn, the first modeling framework with synaptic heterogeneity in SNNs that replaces the conventional membrane time constant with synapse-specific time constants. This design enables versatile timescale integration at the synapse level, grounded in biological observations of synaptic diversity. We instantiate HetSyn as HetSynLIF and theoretically show its generalization ability to existing models. Through comprehensive experiments, we show that HetSynLIF consistently achieves high accuracy on tasks requiring versatile timescale integration, exhibits strong robustness to noise, and maintains computational efficiency under resource constraints. Furthermore, the learned synaptic time constants exhibit a mildly long-tailed distribution, loosely resembling biological observations and reflecting adaptation to multi-timescale processing. These findings underscore the critical role of synaptic heterogeneity in SNNs, not only in enhancing computational performance, but also in aligning with biological principles. By establishing synapse-level modeling as a viable direction, our work points toward promising avenues for advancing brain-inspired learning systems. Future work could explore the combination of HetSyn with other advanced spiking architectures or training strategies for further improvements in versatile timescale integration and task performance.




\bibliography{main}  

\newpage
\appendix

\renewcommand{\thefigure}{A\arabic{figure}} 
\setcounter{figure}{0}
\section*{Appendix}

\section{Theoretical Proof}
\label{Theoretical Proof}

\paragraph{Proposition 1} {\itshape If all synaptic decay factors \(r_{ji}\) and the reset current decay factor \(\kappa_j\) are identical and equal to a shared value \(\rho\), i.e., \(r_{ji} = \kappa_j = \rho\) for all \(i, j\), then the HetSynLIF model reduces to the HomNeuLIF model.}

\paragraph{\textit{Proof}} From Eq.~(\ref{eq:HetSyn_membrane}) and Eq.~(\ref{eq:HetSyn_current}--\ref{eq:HetSyn_spike}), the membrane potential \( V_j^t\) of postsynaptic neuron \(j\) at time \(t\) in the HetSynLIF model is given by:
\begin{align}
V_j^t &= \left( \sum_i r_{ji} \cdot I_{ji}^{t-1} + \sum_i w_{ji} \cdot z_i^t \right) 
- \left( \kappa_j \cdot J_j^{t-1} + \vartheta \cdot z_j^{t-1} \right) \label{eq:proof11}
\end{align}
Under the condition that \( r_{ji} = \kappa_j = \rho \), Eq.~(\ref{eq:proof11}) simplifies to:
\begin{align}
V_j^t &= \left( \sum_i \rho \cdot I_{ji}^{t-1} + \sum_i w_{ji} \cdot z_i^t \right) 
- \left( \rho \cdot J_j^{t-1} + \vartheta \cdot z_j^{t-1} \right) \nonumber \\
&= \rho \cdot \left( \sum_i I_{ji}^{t-1} - J_j^{t-1} \right) 
+ \sum_i w_{ji} \cdot z_i^t - \vartheta \cdot z_j^{t-1} \label{eq:proof12}
\end{align}
Observe that the term \( \sum_i I_{ji}^{t-1} - J_j^{t-1} \) corresponds precisely to the membrane potential at the previous time step \( t - 1\), as defined in Eq.~(\ref{eq:HetSyn_current}):
\begin{align}
V_j^{t-1} = \sum_i I_{ji}^{t-1} - J_j^{t-1} \label{eq:proof13}
\end{align}
Substituting Eq.~(\ref{eq:proof12}) yields:
\begin{align}
V_j^t = \rho \cdot V_j^{t-1} + \sum_i w_{ji} \cdot z_i^t - \vartheta \cdot z_j^{t-1} \label{eq:proof14}
\end{align}

This recurrence relation is identical to the membrane update rule of the HomNeuLIF model, 
as defined in Eq.~(\ref{eq:LIF_membrane_discrete}). Therefore, under the condition \( r_{ji} = \kappa_j = \rho \), the HetSynLIF model reduces exactly to the HomNeuLIF model under the specified condition. This completes the proof.

\paragraph{Proposition 2} \textit{Based on Proposition~1, if we further decompose the reset current into a standard component and an additional adaptation current triggered by spikes—characterized by a decay factor \( \rho_{a} \) and a scaling coefficient \( a \)—then HetSynLIF generalizes to the HomNeuALIF model.}

\paragraph{\textit{Proof}} Under the condition that the total reset current in the HetSynLIF model consists of two components: a standard reset current \( J_{\vartheta}^ t\) and an additional adaptation current \(J_a^t\), such that the membrane potential \(V_j^t\) of postsynaptic neuron \(j\) at time \(t\) satisfies:
\begin{align}
V_j^{t} = \sum_i I_{ji}^{t} - J_{\vartheta}^{t}-J_{a}^{t} \label{eq:proof21}
\end{align}
The standard reset current \(J_{\vartheta}^{t}\) follows the dynamics defined in Eq.~(\ref{eq:HetSyn_spike}), while the adaptation current \(J_{a}^{t}\) follows the dynamics introduced in \cite{bellec2018long}:
\begin{align}
\frac{d J_{a}^{t}}{d t} &=-\frac{J_{a}^{t}}{\tau_{a}}+\sum_{j} a \cdot \delta\left(t-t_{\text{s}}^{j}\right) 
\label{eq:proof22} \\
J_{a}^{t}&=\rho_{a} \cdot J_{a}^{t-1}+a \cdot z_{j}^{t-1}  \label{eq:proof23}
\end{align}
where \(\rho_{a}=\text{exp}(-\frac{\Delta t}{\tau_{a}})\) is the decay factor derived from the adaptation time constant \(\tau_a\), and \(a\) is the adaptation strength. Substituting Eq.~(\ref{eq:HetSyn_current}--\ref{eq:HetSyn_spike}) into Eq.~(\ref{eq:proof21}), we obtain:
\begin{align}
V_{j}^{t} = \sum_{i} \left(r_{ji} \cdot I_{i}^{t-1}+w_{ji} \cdot z_{i}^{t}\right) - \kappa_{j} \cdot J_{\vartheta}^{t-1} -\vartheta \cdot z_{j}^{t-1}-J_{a}^{t}  \label{eq:proof24}
\end{align}
Under the condition that \(r_{ji} = \kappa_{j} = \rho\), Eq.~(\ref{eq:proof24}) simplifies to:
\begin{align}
V_{j}^{t} &= \rho \cdot \left( \sum_i I_{ji}^{t-1} - J_{\vartheta}^{t-1} \right) 
+ \sum_i w_{ji} \cdot z_i^t - J_{a}^{t} -\vartheta \cdot z_{j}^{t-1} \nonumber \\
&=\rho \cdot V_j^{t-1} + \sum_i w_{ji} \cdot z_i^t - J_{a}^{t}-\vartheta \cdot z_{j}^{t-1} \label{eq:proof25}
\end{align}
The recurrence in Eq.~(\ref{eq:proof25}) is similar as the membrane potential dynamics of the HomNeuALIF model defined in \cite{rao2022long}. This completes the proof.

\paragraph{Proposition 3} \textit{For each postsynaptic neuron \(j\), if all synaptic decay factors \(r_{ji}\) and the reset current decay factor \(\kappa_j\) are identical and equal to a neuron-specific membrane decay factor \(\rho_j\), i.e., \(r_{ji} = \kappa_j = \rho_j\) for all \(i\), then HetSynLIF generalizes to the HetNeuLIF model.} 

\paragraph{\textit{Proof}} From Eq.~(\ref{eq:HetSyn_membrane}) and Eq.~(\ref{eq:HetSyn_current}--\ref{eq:HetSyn_spike}), the membrane potential \(V_j^t\) of postsynaptic neuron \(j\) in the HetSynLIF model is given by Eq.~(\ref{eq:proof11}). Under the condition that \(r_{ji} = \kappa_j = \rho_j\), substituting this into Eq.~(\ref{eq:proof11}), we obtain:
\begin{align}
V_j^t = \left( \sum_i \rho_j \cdot I_{ji}^{t-1} + \sum_i w_{ji} \cdot z_i^t \right) 
- \left( \rho_j \cdot J_j^{t-1} + \vartheta \cdot z_j^{t-1} \right) \label{eq:proof31}
\end{align}
Rewriting Eq.~(\ref{eq:proof31}) by extracting \( \rho_j \) as a common factor, we have:
\begin{align}
V_j^t &= \rho_j \cdot \left( \sum_i I_{ji}^{t-1} - J_j^{t-1} \right) 
+ \sum_i w_{ji} \cdot z_i^t - \vartheta \cdot z_j^{t-1} \label{eq:proof32}
\end{align}
Substituting Eq.~(\ref{eq:proof13}) into Eq.~(\ref{eq:proof32}) yields:
\begin{align}
V_j^t = \rho_j \cdot V_j^{t-1} + \sum_i w_{ji} \cdot z_i^t - \vartheta \cdot z_j^{t-1} \label{eq:proof33}
\end{align}
This update rule corresponds to the membrane dynamics of the HetNeuLIF model, in which each neuron has its own membrane decay factor \(\rho_j\). Thus, under given condition, the HetSynLIF model reduces to the HetNeuLIF model. This completes the proof. 

\section{Details of the Pattern Generation Task}
\label{Details in Patter Generation task}


\paragraph{Task Description} In this task, network is trained to reproduce a continuous target trace in response to structured input spike patterns. Each input consists of 100 afferent channels spanning a 2000-ms window. A single 500-ms spike segment is first generated using a 10 Hz Poisson process and then embedded three times within the input window, evenly spaced and separated by 250-ms silent intervals (i.e., periods with no input spikes from the structured segment, though background noise may still occur). To perturb the input structure, global background noise is superimposed across all channels, and regenerated independently at each training iteration. In our experiments (Fig.~\ref{fig:fig2}E, F), we evaluate noise levels with rates of 1, 2, 3, and 4 Hz to examine the model's robustness under increasing perturbation. The target trace is constructed as the sum of five cosine components with a base frequency of 0.5 Hz, with each component modulated by a random amplitude sampled from \(\mathcal{U}(0,1)\) and a phase sampled from \(\mathcal{U}(0, 2\pi)\). The resulting signal is then normalized to the range \( [-1, 1]\).

\paragraph{Loss Function} We train the network using a loss that combines a task-specific term and a regularization term. The task loss is defined as the mean squared error (MSE) between the network output (\( y^t \)) and the target trace (\( y^{*, t}\)), averaged over all timesteps. The regularization term penalizes deviations of each neuron's firing rate from a target rate of 10 Hz, computed as the mean squared relative error across neurons. The two terms are weighted using a fixed coefficient of 0.01 for the regularization.

\paragraph{Computation of the Normalized Distances} In Fig.~\ref{fig:fig2}B and C, we quantify the model’s ability to generate temporally distinct outputs for repeated input segments by computing the normalized distance between the output traces corresponding to different occurrences of the same structured input. Specifically, for any two repeated segments (e.g., the first and second), we compute the mean squared difference between their output traces, and normalize it by the corresponding difference between the target traces:

\begin{align}
\text{dist}(\text{2nd},\text{1st}) = \frac{\sum_{t}\left(y_{\text{2nd}}^{t}-y_{\text{1st}}^{t}\right)^2}{\sum_{t}\left(y_{\text{2nd}}^{*,t}-y_{\text{1st}}^{*,t}\right)^2}
\end{align}

Here, a higher distance indicates that the model produces differentiated outputs for repeated inputs, suggesting alignment with the target trace rather than generating identical responses to identical stimuli. Conversely, a lower value implies that the model outputs remain nearly the same across repetitions.

\paragraph{Experiment Settings} We adopt a 100-100-1 network architecture for all models, comprising 100 input channels, 100 hidden neurons, and a single leaky-integrate readout neuron. The synaptic time constant \(\tau_{\text{s}}\) is fixed at 20~ms for HomNeuLIF and HomNeuALIF, and initialized by sampling from \(\mathcal{N}(20, 5)\)~ms for HetNeuLIF and HetSynLIF. The reset time constant \(\tau_J\) is uniformly set to 20~ms across all models. For HomNeuALIF, the adaptation time constant and adaptation strength are set to \(\tau_{a} = 500\)~ms and \(a = 0.01\), respectively. All decay factors are constrained to the range \([0, 1]\) during training. For the surrogate function, we use a triangular-shaped derivative defined as \( \frac{\partial z^t}{\partial V^t}=\gamma \cdot \max \left(0,1-\left|\frac{V^{t}-\vartheta}{\zeta \cdot \vartheta}\right|\right) \), we set \(\gamma=1\) and \(\zeta=1\). Training is performed using the Adam optimizer with a StepLR learning rate scheduler, where the learning rate is initialized at 1e-3 and decayed by a factor of 0.8 every 100 iterations. All models are trained for 1000 iterations with a batch size of 1.


\section{Details of the Delayed Match-to-Sample Task}
\label{Details in Delayed Match-to-Sample Task}

\paragraph{Task Description}  
In this task, the network is required to determine whether two temporally separated cues belong to the same category (i.e., left or right). There are four possible cue combinations: \texttt{left-left}, \texttt{left-right}, \texttt{right-left}, and \texttt{right-right}. The \texttt{left-left} and \texttt{right-right} conditions are labeled as “match” , while the remaining two are considered “mismatch”. The input consists of 30 channels, divided equally into left-cue, right-cue, and noise channels (10 each). Each cue is delivered through either the left or right group using independent Poisson spike trains at 40~Hz. The first cue begins at 50~ms and lasts for 100~ms, followed by an 800~ms delay. The second cue then appears, also lasting 100~ms. Throughout the entire 1050~ms input window, continuous background noise is added on the noise channels using a 10~Hz Poisson process. To evaluate the model’s ability to retain the first cue and compare it to the second after a long delay, we compute the decision by comparing the membrane potentials of two output neurons—corresponding to “match” and “mismatch”—at the final millisecond of the second cue (1050~ms), and the class associated with the higher membrane potential is selected as the predicted label.

\paragraph{Loss Function}
Similar to the pattern generation task, the network is trained using a loss function composed of a task-specific term and a regularization term. The regularization term remains unchanged and penalizes deviations of each neuron's firing rates from a target of 10~Hz. The task-specific loss is defined as the cross-entropy between the predicted class probabilities and the target labels.

\paragraph{Experiment Settings}
In this task, the synaptic time constant \(\tau_{\text{s}}\) is fixed at 40~ms for HomNeuLIF and HomNeuALIF, and initialized by sampling from \(\mathcal{N}(40, 4)\)~ms for HetNeuLIF and HetSynLIF. The reset time constant \(\tau_J\) is uniformly set to 40~ms across all models. For HomNeuALIF, the adaptation time constant and adaptation strength are set to \(\tau_{a} = 800\)~ms and \(a = 0.15\), respectively. All decay factors are constrained to the range \([0, 1]\) during training. A simulation timestep of \( \Delta t = 5 \)~ms is used to reduce computational overhead. We use the same surrogate function as in the pattern generation task and set \(\gamma=1\) and \(\zeta=1\). Training is performed using the Adam optimizer with a StepLR learning rate scheduler, where the learning rate is initialized at 5e-3 and decayed by a factor of 0.8 every 100 iterations. All models are trained for 1000 iterations with a batch size of 32.


\section{Details of the Speech Recognition Task}
\label{Details in Speech Recognition Task}

\paragraph{SHD Dataset and Preprocessing}
The SHD dataset comprises 10,420 high-quality audio samples of spoken digits (0-9) in English and German. It includes 12 speakers—6 female and 6 male—aged between 21 and 56, with each speaker contributing approximately 40 utterances per digit for each language. The dataset is divided into training and testing sets, containing 8,156 and 2,264 samples, respectively. Before feeding into neural networks, we first align all audio recordings to a fixed duration of 1000 ms by trimming or zero-padding, and then sample the resulting spike trains using a 4 ms time bin, yielding a 250 $\times$ 700 input matrix per recording (250 timesteps $\times$ 700 input channels).

\paragraph{Deletion Noise}
Spiking systems deployed in real-world scenarios often face partial signal loss or sensor failures, making robustness to missing inputs a critical property. To assess this capability, we evaluate performance under deletion noise, which simulates missing input events by independently removing each spike with a fixed probability \( p \). Specifically, for each spike input (i.e., binary value 1), it is retained with probability \( 1 - p \) and set to 0 with probability \( p \). We test seven noise levels with \( p \in \{ 0.0, 0.01, 0.02, 0.04, 0.08, 0.16, 0.32, 0.64 \} \), corresponding to \( p = 0.01 \times 2^n \) for \( n = 0 \) to \( 6 \), plus a clean condition. 

\paragraph{Temporal scaling}
Temporal scaling is a fundamental aspect of sensory processing, as biological systems such as the auditory and motor systems are capable of recognizing patterns across a wide range of speeds. To test whether SNNs equipped with HetSyn exhibit similar robustness, we simulate temporal variability by scaling each spike time with a global warping factor. That is, each spike time is multiplied by a constant factor \( \alpha \), resulting in a globally compressed or stretched spike sequence. Accordingly, the total number of timesteps changes proportionally to the warping factor. We evaluate 20 different warping conditions, with factors sampled as \( \log_2 \alpha \sim \mathcal{U}(-0.5, 0.5)\), covering both temporal compression (\( \log_2 \alpha < 0 \)) and dilation (\( \log_2 \alpha > 0 \)) at varying scales.

\paragraph{Experiment Settings}
As in \cite{yin2021accurate, zheng2024temporal}, we adopt leaky-integrate neurons for the readout layer to decode the output of the network, where the predicted possibility for class $i$ is obtained by summing the softmax-normalized membrane potential of the output neuron over time, i.e., $\hat{y}_i = \sum_t\text{softmax} (V_{\text{out}}^t)$. For better performance, we use multi-Gaussian curve\cite{yin2021accurate} as surrogate function, and train the network using the standard cross-entropy loss. Besides, a dropout rate of 0.2 is applied to HetSynLIF, HetNeuLIF, and HomNeuLIF, and 0.3 for HetNeuALIF based on its greater tendency to overfit observed during training. The synaptic time constant \(\tau_{\text{s}}\) is fixed at 20~ms for HomNeuLIF and HomNeuALIF, and initialized by sampling from \(\mathcal{U}(5, 20)\)~ms for HetNeuLIF and HetSynLIF. The reset time constant \(\tau_J\) is uniformly set to 20~ms across all models. For HomNeuALIF, the adaptation time constant and adaptation strength are set to \(\tau_{a} = 100\)~ms and \(a = 0.05\), respectively. All decay factors are constrained to the range \([0, 1]\) during training. The model is optimized using the AdamW with a learning rate of 2e-3 and a weight decay of 4e-3. We apply a cosine annealing schedule with 5\% warm-up, computed per batch, and fix the learning rate after 40 epochs.


\section{Datasets and Configurations}
\label{Details in Comparison with Existing Works}

\paragraph{SHD} Same as Appendix \ref{Details in Speech Recognition Task}.

\paragraph{S-MNIST}  
Sequential MNIST (S-MNIST) is a sequential version of the standard MNIST dataset, where each \(28 \times 28\) image is reshaped into a sequence of 784 inputs. At each timestep, a single pixel is presented to the model in a row-wise scan from top-left to bottom-right. The MNIST dataset consists of grayscale images of handwritten digits (0–9), with 60,000 training and 10,000 test samples. We follow the original train/test split in our evaluation.

\paragraph{TiDigits} We use the adults subset of the TiDigits dataset, which comprises isolated utterances of the 11 English digit classes (“zero”–“nine” and “oh”), with standard training and testing splits of 2,464 and 2,486 speech samples. We use threshold crossing encoding method proposed by Gutig \cite{gutig2016spiking} to encode TiDigits into spikes, with nfft=512, 16 Mel filters from 360–8000 Hz, and 15 thresholds per channel for crossing. We use the first 200 timesteps of each sample, yielding an input of shape 200 \( \times \) 500 (timesteps \( \times \) channels).

\paragraph{Ti46-Alpha} Ti46-Alpha, consisting of the 26 English alphabet letters, is a subset of the Ti46-Word dataset. We apply the same encoding method as in TiDigits, yielding input representations of size 200 \( \times \) 500.

\paragraph{Experiment settings} All experiments are conducted using NVIDIA RTX 4090 and Tesla V100-PCIE-16GB GPUs. We employed a cosine-annealing learning rate scheduler applied at each batch step. For S-MNIST, classification is based on spike counts from the output layer neurons, whereas for all other datasets, predictions are derived from temporally integrated outputs, as previously described for SHD. A detailed overview of the experimental settings and hyperparameter configurations is provided in Table \ref{table:hyperparameter_config}.

\begin{table}[htbp]
    \centering
    \caption{Experiment settings and hyperparameter configurations for different datasets}
    \vspace{\baselineskip}
    \label{table:hyperparameter_config}
    \begin{tabular}{lllll}
    \toprule
        Dataset&SHD&S-MNIST&TiDigits&Ti46-Alpha\\
        \cmidrule(r){1-5}
        learning rate&2e-3&1e-3&2e-3&4e-3\\
        \cmidrule(r){1-5}
        dropout rate&0.2&0&0.5&0.2\\
        \cmidrule(r){1-5}
        epochs&100&150&50&50\\
        \cmidrule(r){1-5}
        batch size&32&256&32&32\\
        \cmidrule(r){1-5}
        warmup ratio&0.05&0.05&0.05&0.05\\
        \cmidrule(r){1-5}
        optimizer&AdamW&AdamW&AdamW&AdamW\\
        \cmidrule(r){1-5}
        weight decay&4e-3&0&1e-2&4e-3\\
        \cmidrule(r){1-5}
        architecture&SRNN&SRNN&SRNN&SRNN\\
        \cmidrule(r){1-5}
        hidden neuron number&[128, 64]&[64, 64]&[64, 32]&[128, 128]\\
        \cmidrule(r){1-5}
        $\vartheta$&1.0&1.0&1.0&1.0\\
        \cmidrule(r){1-5}
        $\Delta t$&1e-3&1e-3&1e-3&1e-3\\
        \cmidrule(r){1-5}
        $\tau_J$&20e-3&20e-3&20e-3&20e-3\\
        \cmidrule(r){1-5}
        initialization of $\tau_\text{s}$&\(\mathcal{U}\)(5e-3, 20e-3)&\(\mathcal{U}\)(5e-3, 20e-3)&\(\mathcal{U}\)(5e-3, 20e-3)&\(\mathcal{U}\)(5e-3, 20e-3)\\
        
    \bottomrule
    \end{tabular}
\end{table}

\section{Discussion}
\label{Discussion}

\paragraph{Broader impacts} By introducing synaptic heterogeneity into SNNs, our work advances the modeling of biologically inspired temporal dynamics and enhances the computational performance of SNNs. This has the potential to improve both learning performance and energy efficiency in neuromorphic computing systems, contributing to the development of low-power, event-driven AI applications. Furthermore, our study is purely computational and does not involve sensitive data or human subjects, and we anticipate no adverse societal or environmental impacts of out study.


\paragraph{Limitations} This work provides a step toward understanding the role of synaptic heterogeneity in SNNs by examining a subset of synaptic properties. A wide range of other synaptic features and more sophisticated architectures remains to be explored in future work.
\newpage

\section{Supplementary Figures}

\vspace*{\fill}

\begin{center}
  \includegraphics[width=1.0\textwidth]{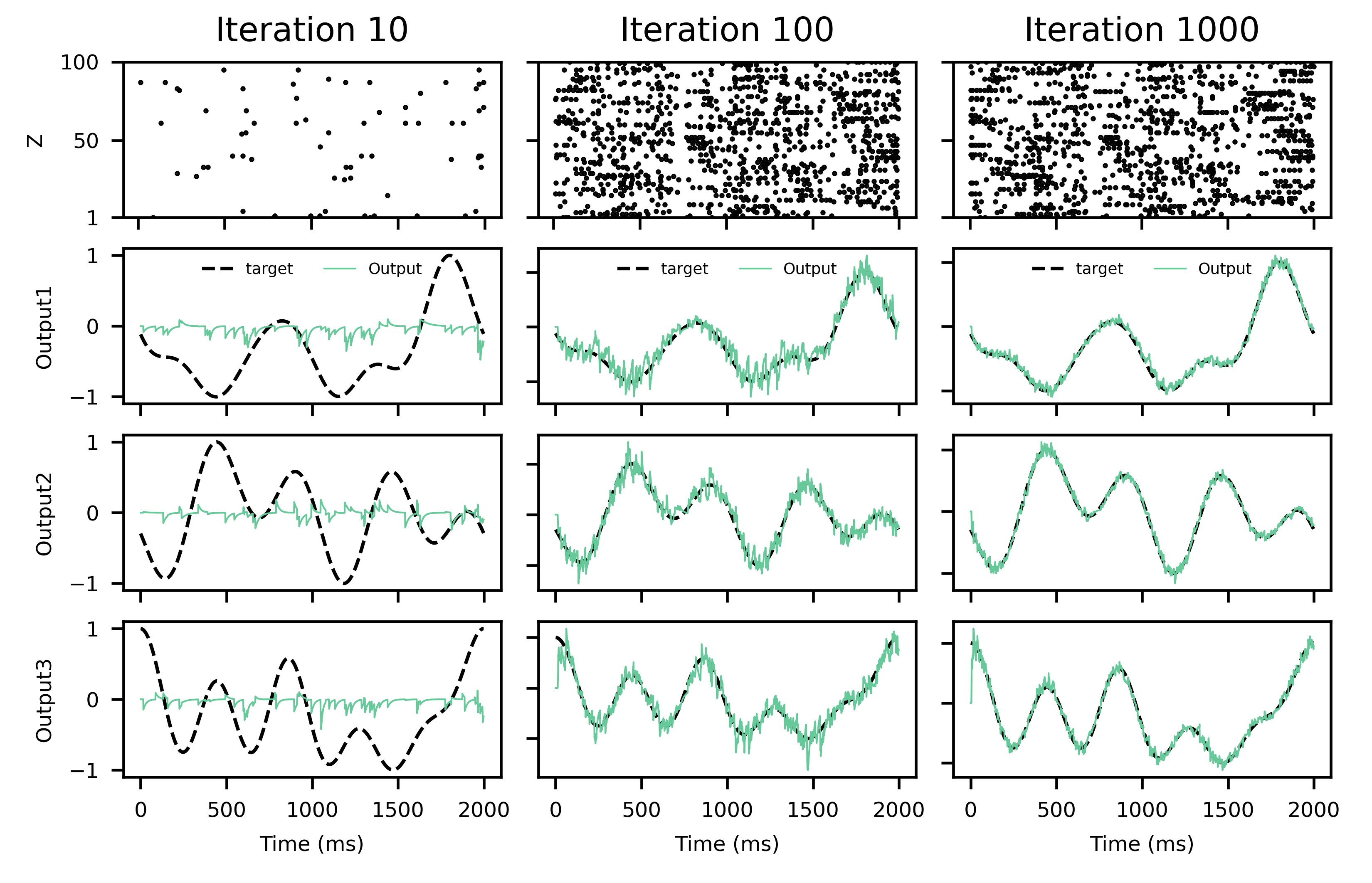}
  \captionof{figure}{Dynamics of F-HetSynLIF during training on the pattern generation task with simultaneous generation of three output patterns. From left to right: network dynamics at 10, 100, and 1000 training iterations. Each column shows spike activity of hidden neurons (top), followed by predicted traces (colored lines) and targets (black dashed lines) for the three outputs.}
  \label{fig:fig5}
\end{center}

\vspace*{\fill}
\clearpage

\vspace*{\fill}

\begin{center}
  \includegraphics[width=1.0\textwidth]{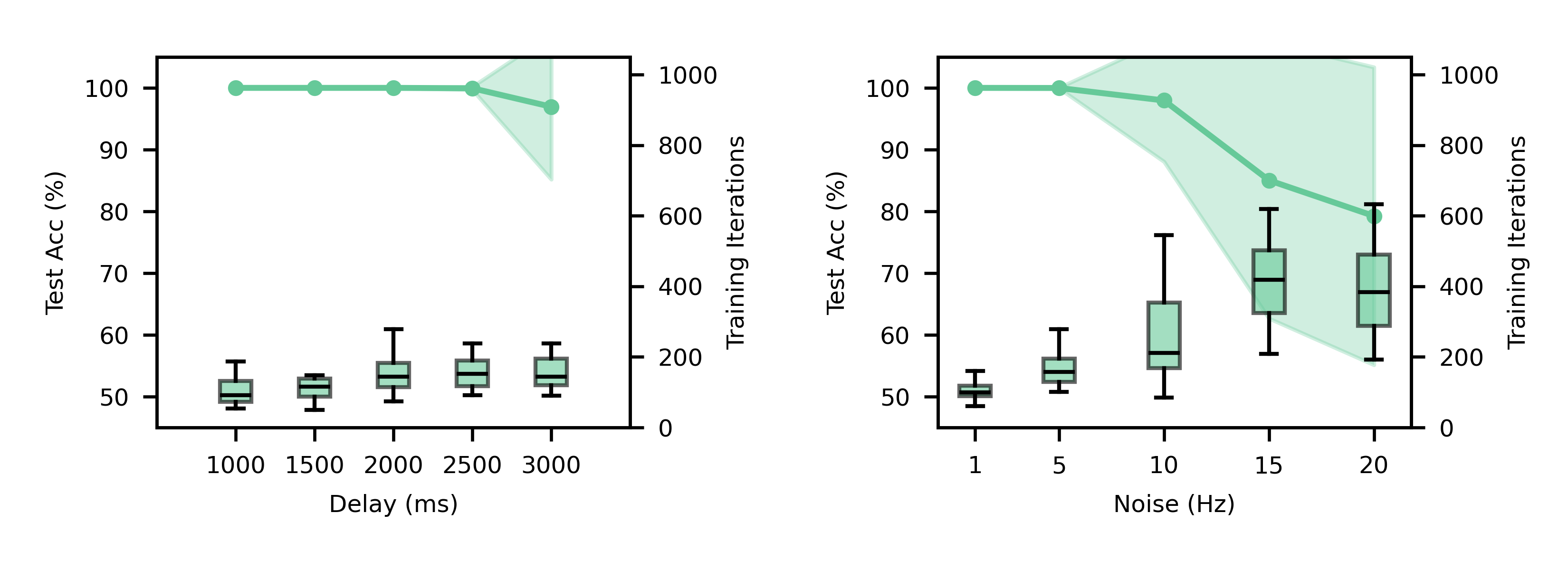}
  \captionof{figure}{Performance and training efficiency of R-HetSynLIF on the delayed match-to-sample task under varying delays and noise levels. Left: Test accuracy (solid line, left y-axis) and training iterations required to reach 95\% accuracy (boxplot, right y-axis; window size = 10) across different delay durations. Right: Same metrics evaluated under varying levels of addition noise (Hz).}
  \label{fig:fig6}
\end{center}

\vspace*{\fill}
\clearpage

\vspace*{\fill}

\begin{center}
  \includegraphics[width=1.0\textwidth]{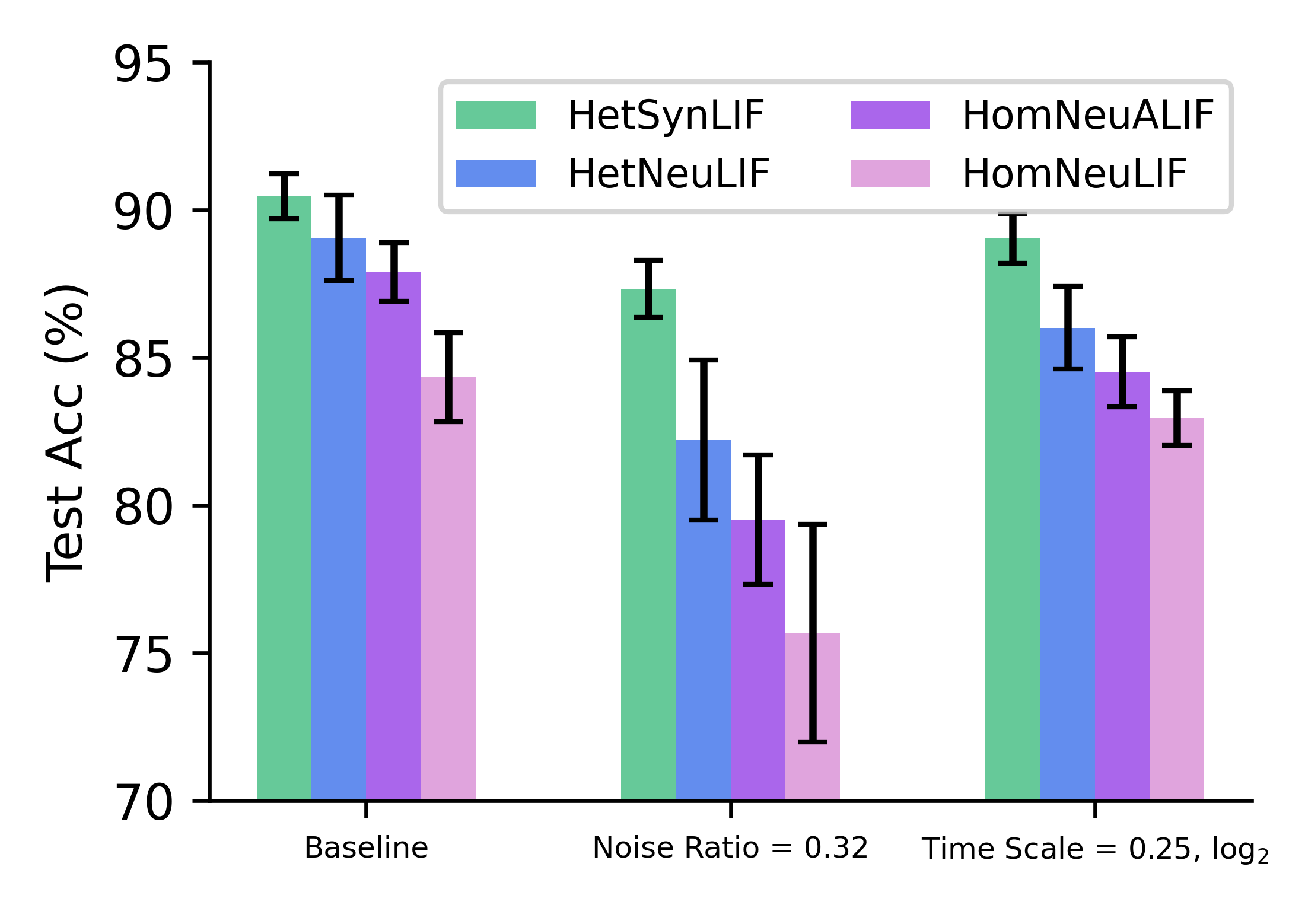}
  \captionof{figure}{Bar plots show test accuracy of four RSNN variants under three conditions: baseline (left), deletion noise ratio 0.32 (middle, from Fig.~\ref{fig:fig4}C), and time warp scale 0.25 ($\text{log}_2$, right, from Fig.~\ref{fig:fig4}F). Error bars represent standard deviation over 10 runs.}
  \label{fig:fig7}
  
\end{center}

\vspace*{\fill}
\clearpage

\vspace*{\fill}
\begin{center}

  \includegraphics[width=1.0\textwidth]{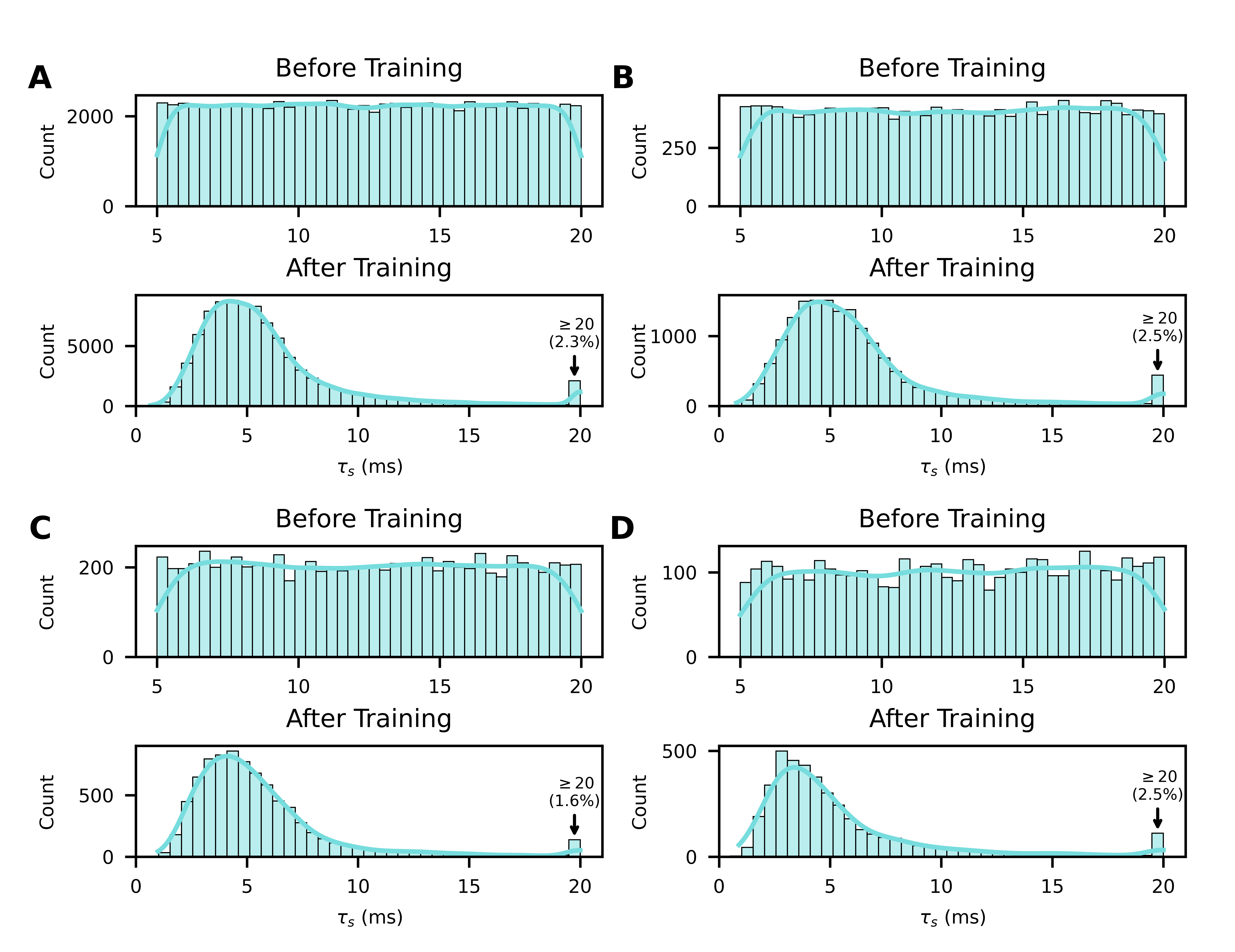}
  \captionof{figure}{Distributions of synaptic time constants ($\tau_\text{s}$) across all connection types in the two-layer R-HetSynLIF. From (A) to (D): input to hidden layer 1, recurrent connections within hidden layer 1, hidden layer 1 to hidden layer 2, and recurrent connections within hidden layer 2. Each subplot shows the distribution of $\tau_\text{s}$ before training (top) and after training (bottom).}
  \label{fig:fig8}

\end{center}
\vspace*{\fill}

\clearpage


\end{document}